# MAXIMUM ALLOWABLE TEMPERATURE DURING QUENCH IN NB3SN ACCELERATOR MAGNETS

G. Ambrosio, Fermilab, Batavia, IL 60510, USA

*Abstract*

This note aims at understanding the maximum allowable temperature at the hot spot during a quench in $Nb_3Sn$ accelerator magnets, through the analysis of experimental results previously presented.

## INTRODUCTION

$Nb_3Sn$ accelerator magnets under development for possible use in the Large Hadron Collider [1,2] may reach, during a quench, higher hot spot temperatures than presently-used Nb-Ti accelerator magnets. This is due both to the higher critical current density in the non-copper section and to the lower copper-non-copper ratio in $Nb_3Sn$ strands than in Nb-Ti strands, together with their different cooling properties. Therefore, understanding the maximum allowable hot spot temperature in $Nb_3Sn$ accelerator magnets has primary importance in the design of these magnets and their protection systems.

In this report this question is addressed through the analysis of tests previously performed on a quadrupole, on a small racetrack, and on some cable samples made with internal tin $Nb_3Sn$ strands.

## HIGH TEMPERATURE TESTS ON A NB$_3$SN QUADRUPOLE

The quadrupole which was the subject of the test discussed here is TQS01: the first Technological Quadrupole with shell structure assembled by LARP [3]. This 1-m-long, 90-mm-aperture magnet was assembled and cold tested three times. At the end of the last test (TQS01c) [4], performed at Fermilab in 2007, high hot spot temperatures were reached in order to evaluate their impact on the magnet's performance. This experiment was performed at 4.6 K bath temperature and the magnet was operating at about 80% of the short sample limit when the experiment started. TQS01c used a Modified Jelly Roll (MJR) conductor manufactured by Oxford Superconducting Technology (OST) with 47% copper. Since TQS01c had no operating spot heaters at the time of this test, spontaneous quenches were used. All spontaneous quenches during this experiment occurred in the same segment (very likely in the same location) in the pole turn of the inner layer of a single coil.

High hot spot temperatures were reached by increasing the delays of dump resistor and protection heaters before the High Temperature (HT) quenches (diamond and triangular markers in Figs. 1 and 2). Increased hot spot temperatures could be reached by increasing these delays. During the experiment some standard quenches (square markers in Figs. 1 and 2) were performed in order to access magnet performance reproducibility and possible detraining effects.

Figure 1 shows that the test started with current ramps to quench at 250 A/s (diamond markers), after which no degradation was found (first four square markers). Subsequently the ramp rate was decreased to 20 A/s in order to reach higher currents and temperatures. Then after five HT quenches (triangular markers) with negligible effects, the 6$^{th}$ HT quench caused an increase of the quench current by 3.3%. The subsequent HT quench caused a detraining of 7.2% with respect to the quench current previously reached. The detraining was recovered after one standard quench, and the subsequent standard quenches confirmed the gain achieved after the 6$^{th}$ HT quench. The 8$^{th}$ HT quench caused a small detraining after which the magnet reached the highest quench current during the entire experiment (4% higher than the quench current plateau before starting the HT experiment). In the subsequent HT quenches at higher and higher temperatures TQS01c showed more and more degradation. Standard quenches showed some permanent degradation after the 14$^{th}$ and 15$^{th}$ HT quenches. At the end of the experiment the permanent degradation was about 25% with respect to the quench current at the beginning of the experiment.

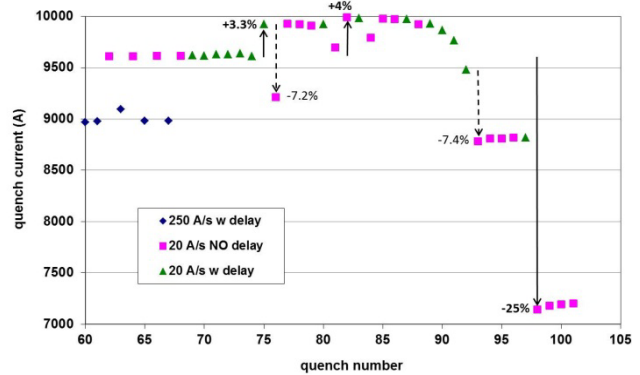

Figure 1: Quench history during high hot spot temperature experiment performed at the end of TQS01c test. Triangular markers show high temperature quenches with long protection delays. Square markers show standard quenches.

The hot spot temperature could not be measured because of the lack of dedicated instrumentation. Therefore the temperature was computed from the measured values of the quench integral (integral of current squared vs. time from the quench start). The code QuenchPro [5] was used to do this computation under the following assumptions:

- Adiabatic approximation.
- The following components were taken into account in the computation of the peak temperature from the

quench integral: the metals in the Rutherford cable, the epoxy within the cable, and the cable insulation (0.1 mm thick assuming some compression after heat treatment). The resulting material fractions are: $Nb_3Sn$ = 23.7%; Cu = 31.5%; bronze = 11.7%; G10 = 33.2%.

- In QuenchPro the copper properties depend on the temperature and on the Residual Resistivity Ratio (RRR), whereas the field is assumed to be constant. In this analysis the cable peak field was used.
- The RRR was measured during magnet test, but the RRR of the quenching segment was not available. Therefore the analysis was performed for the max and min RRR values (170-130) of the quenching coil. The impact of this uncertainty is +/- 6 K with respect to the values shown in Fig. 2.

The results of the hot spot temperature computation are shown in Fig. 2. This is the same quench history plot shown in Fig. 1 with the hot spot temperature reached in most HT quenches. The temperatures (in K) shown on the plot were computed using the average RRR of the quenching coil.

Figure 2 shows that: i) quenches with temperature in the hot spot ($T_{HS}$) around 340 K caused very small quench current changes; ii) quenches with 370 K < $T_{HS}$ < 400 K caused reversible current changes of a few per cent; iii) quenches with $T_{HS}$ > 460 K caused irreversible degradation.

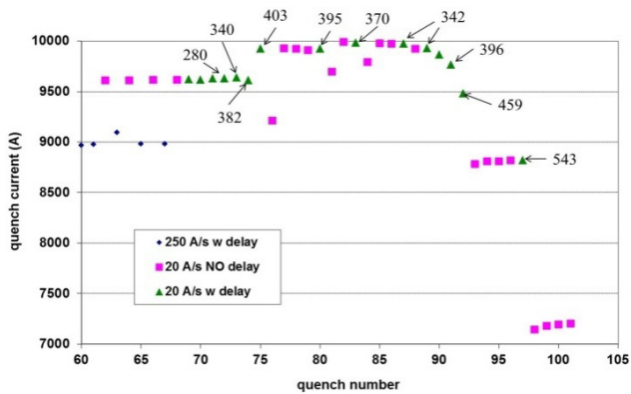

Figure 2: Quench history during high hot spot temperature experiment performed at the end of TQS01c test. The numbers show the peak temperature (in K) reached at the hot spot in some HT quenches.

## TESTS PERFORMED ON A $NB_3SN$ SMALL RACETRACK AND CABLE SAMPLES

A useful set of test results and analysis is presented in [6]. High temperature quenches were performed on cables at the NHMFL and on a small racetrack magnet at LBNL. The cables were made of 0.7 mm-diameter ITER-type strands manufactured by IGC Advanced Superconductors with 59% copper fraction. Two samples (Cable 2-a and 2-b) had bending strain induced after reaction; the other sample (Cable 1) did not have any bending strain. The small racetrack (SM05) was made of two coils. The coil used for the high-temperature quenches was instrumented with a spot heater and voltage taps close to the spot heater. This coil was made of MJR strands manufactured by OST with 0.67 mm diameter and 60% copper fraction.

The test results are presented in Fig. 3 (from Ref. [6]). The horizontal axis shows the peak temperature reached in each HT quench. The vertical axis shows the reduced current (quench current divided by maximum current) reached in the standard ramp to quench following each HT quench. Therefore each point shows the degradation vs. hot spot temperature. All cables and the racetrack magnet were instrumented with spot heaters for initiating the quench and with voltage taps around the hot spot area. The resistance growth measured by these voltage taps was used to compute the peak temperature, providing a precise although indirect measurement. A comparison between these measurements and computations using the quench integral is presented in the following section.

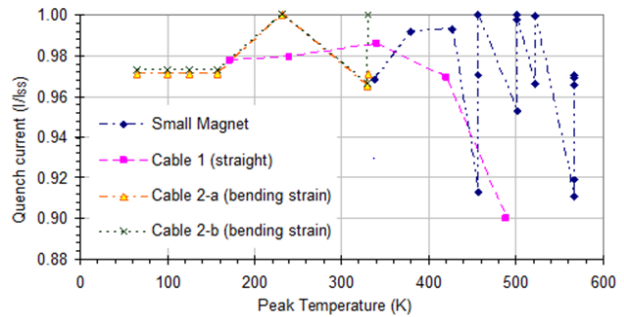

Figure 3: Summary of quench experiments: reduced current vs. peak temperature reached during the preceding HT quench test. The lines represent the temporary sequence of the high temperature events (from Ref [6]).

The plot in Fig. 3 shows negligible degradation up to 420 K. At higher temperatures the small racetrack started detraining and retraining between 90% and 100% of the short sample limit and reached about 570 K with a degradation of only 3%. The cable sample 1, after a HT quench at ~480 K, showed a degradation of 8% together with an insulation failure that irreversibly damaged the sample. This failure demonstrates that the maximum allowable temperature does not depend only on critical current degradation, but also on insulation integrity.

Ref. [6] also presents an interesting comparison between simulations and experimental data collected during a series of cable quench tests. Figure 4 shows different computations of the Quench Integral (QI): (i) using only the metals in the Rutherford cable; (ii) adding the epoxy included in the cable envelope; and (iii) adding also the cable insulation (0.1 mm thick fiberglass tape cured with ceramic binder [7] - resulting in 0.15 mm thickness - and impregnated with epoxy) that was simulated using G10 material properties. Figure 4 also shows the experimental values of the quench integral (square markers with internal cross) in different quenches. The experimental temperature was measured by the resistance growth of the short segment under the spot heater.

It can be seen that when the peak temperature was about 140 K, the QI computed using metal and epoxy was in good agreement with the experimental value. At higher peak temperatures the experimental values approached the QI computed using also the cable insulation. In the 300-400 K range the QI computed including the cable insulation provided the best agreement with the experimental values. Nonetheless it should be noted that including the cable insulation did not provide a conservative estimate in this temperature range.

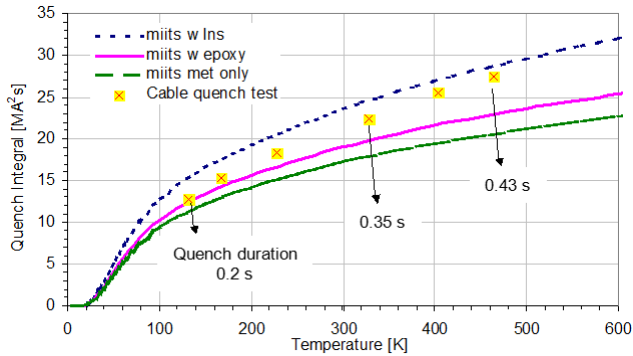

Figure 4: Quench integral of a cable sample vs. temperature: experimental results (square markers) and values computed with different assumptions (dashed line: metals only; continuous line: metals and epoxy inside the cable envelope; dotted line: metals, epoxy and cable insulation). Plot from Ref [6].

## ANALYSIS AND CONCLUSIONS

The set of experimental results presented above suggests some preliminary conclusions, which should be confirmed by further tests.

When the hot spot of a $Nb_3Sn$ accelerator magnet exceeds room temperature, there are two threshold temperatures above which magnet performance may change. We start this analysis by naming these thresholds $T_1$ and $T_2$ and describing the possible effects when the hot spot temperature ($T_{HS}$) exceeds them.

If $T_{HS} > T_1$, then the magnet enters an "active territory" with the following features:
- The magnet may experience further training: i.e. a magnet whose training was completed by reaching a current plateau may actually exceed that current plateau in quenches following a high-temperature quench.
- The magnet may experience detraining: i.e. a reduction of the quench current after a high-temperature quench, which can be recovered with a few training quenches.

If $T_{HS} > T_2$, then the magnet enters a "degradation territory" with the following features:

- The magnet may experience irreversible degradation.
- The magnet may experience insulation degradation with possible failure under stress conditions, for instance during subsequent quenches even at lower hot spot temperatures.

Based on this characterization, the "active territory" appears to be associated with small changes of strain in the conductor (within the reversible region) and small changes of stress in the epoxy, which may cause further training or detraining. The "degradation territory" appears to be associated with larger change of strain in the conductor (above the irreversibility limit) and with large deformations of the epoxy, which may also cause cracks or other degradations of the insulation.

The experimental results presented in Fig. 3 suggest that $T_1$ is around 400 K (disregarding the results of the samples with bending strain, which may have been affected by the special strain condition). The results presented in Fig. 2 (TQS01c) suggest that $T_1$ is between 340 and 370 K, but this estimate may have a large error because the Fig. 2 temperatures were computed whereas the temperatures in Fig. 3 were measured. Estimating the error of the temperatures in Fig. 2 requires a significant effort because it should address both the error due to the material properties used in the computation as well as the error due to each assumption. Figure 4 suggests a different approach. The computed values (dotted line) and the measured values (square markers with a cross) can be used to evaluate the error when the temperature is estimated by taking into account the cable insulation in the quench integral. This comparison shows that the hot spot temperature ($T_{HS}$) would have been underestimated by about 30 K when close to 400 K. The cable insulation used in TQS01c was made of the same materials (fiberglass with ceramic binder impregnated with CTD-101K epoxy) used for the insulation of the cable with test results presented in Fig. 4. The same material properties were used to compute the quench integral used in Fig. 4 (dotted line) and to compute the temperatures in Fig. 2. Therefore we may assume that a similar error should affect both of them. If we apply this correction to the estimate of $T_1$ based on Fig. 2 we obtain: 370 K < $T_1$ < 400 K (with an error that should be no larger than the correction applied, i.e. +/- 30 K).

The quadrupole magnet (TQS01) and the cables with test results presented in Figures 1 to 4 were impregnated using CTD-101K epoxy made by Composite Technology Development (CTD). The small racetrack magnet was impregnated with CTD-101A epoxy made by the same vendor. The glass transition temperature ($T_g$) of CTD-101K is 386 K (113 °C) [8-9]. CTD-101A has thermal and structural properties very similar to those of CTD-101K (for instance its $T_g$ is 388 K) [10]. Above the glass transition temperature the epoxy is in a rubber-like state, which may explain the features previously described when $T_{HS}$ is higher than $T_1$ (active territory). During the high-temperature quenches the hot spot reached temperatures significantly higher than the rest of the coil or cables. The thermal expansion of the hot spot area was larger than the expansion in the rest of the coil or cables, causing significant thermo-mechanical stresses. When the hot spot exceeded $T_g$, the epoxy became soft and susceptible to deformation under the thermo-mechanical stresses. When the temperature decreased below $T_g$, the epoxy returned to its hard state in the new dimensional configuration. For instance, if the hot spot in TQS01c was on the thin edge of a cable in the inner layer, some epoxy could be "extruded" toward the aperture. Signs of this behaviour can be seen in the cross section of the TQS01c quenching coil at the position where all high-temperature quenches initiated [3]. The analysis of TQS01c strain gauges [3] showed a reduction of azimuthal preload in the quenching coil during the high-temperature quenches, confirming that the high-temperature quenches caused epoxy softening and redistribution.

The features associated with the "active territory" can be explained by the redistribution of the epoxy around the hot spot, which may cause a change of strain in the conductor and a change of stress in the epoxy. If $T_{HS}$ slightly exceeds $T_g$, then the epoxy above $T_g$ is limited to a small volume and the possible change of conductor strain remains very likely within the reversible region. If $T_{HS}$ exceeds $T_g$ by a large amount, than the epoxy volume above $T_g$ can be large causing significant changes of conductor strain and possibly irreversible degradation. This analysis suggests that $T_2$, the threshold for the "degradation territory", should be higher than $T_1$. Nonetheless, if the magnet insulation scheme is not sufficiently robust, the thermo-mechanical stresses during a quench (even at moderate hot spot temperatures) could degrade the insulation and lead to electrical failures. Therefore, the insulation scheme of any $Nb_3Sn$ accelerator magnet should be designed to withstand the thermo-mechanical stresses (both within coils and coil-to-structure) well above the glass transition temperature of the epoxy (or other material) used for coil impregnation. By doing so the magnet designers assure that $T_2$ is higher than $T_1$. Since we have demonstrated that $T_1 = T_g$, the glass transition temperature of the epoxy can be used to set the maximum allowable temperature ($T_{max}$) at the magnet hot spot. In order to have some margin $T_{max}$ should be lower than $T_g$. Since we have seen that in a well-designed magnet $T_g$ is not the edge of a cliff, then a 20% margin is sufficient. The margin can be as low as 10% when conservative approximations are used for computing the hot spot temperature, and the error is smaller than the margin.

Therefore, for the design of $Nb_3Sn$ accelerator magnets using CTD-101K epoxy (with $T_g$ = 386 K), we suggest setting the maximum allowable temperature in the hot spot at 350 K or lower. This temperature appears to be consistent with the test results presented in this note and with many tests performed on $Nb_3Sn$ R&D magnets around the world [11].

Finally, it should be noted that none of the magnets and cable samples discussed in this note had a cored cable. The possible impact of a metallic core inside the cable on the maximum allowable temperature during quench should be addressed by a series of dedicated experiments.

## ACKNOWLEDGMENT

Several people contributed to the work presented in this note. Special thanks to Linda Imbasciati, Shlomo Caspi, Guram Chlachidze, Dan Dietderich, Helene Felice and Ezio Todesco.